\documentclass[aps,twocolumn,nofootinbib,preprintnumbers,groupedaddress,letterpaper]{revtex4}

\usepackage[dvipsnames]{xcolor}
\usepackage[normalem]{ulem}
\usepackage{amsmath}
\usepackage{enumerate}
\usepackage{amsfonts}
\usepackage{epsfig}
\usepackage{yfonts}
\usepackage{undertilde}

\usepackage{tikz-cd}

\usepackage{graphicx}
\usepackage{bm}
\usepackage{subfigure}

\newcommand\comment[1]{}

\newcommand\poincare{Poincar\' e }

\newcommand\om{\omega}

\newcommand\ov{\over }

\def\le{\left}
\def\ri{\right}

\def\({\left(}
\def\){\right)}
\def\<{\langle}
\def\>{\rangle}
\newcommand\half{{\ensuremath{\frac{1}{2}}}}
\newcommand\p{\ensuremath{\partial}}

\newcommand\field[1]{{\ensuremath{\mathbb{{#1}}}}}

\newcommand{\RR}{\field{R}}

\newcommand{\be}{\begin{equation}}
\newcommand{\ee}{\end{equation}}
\newcommand{\bea}{\begin{eqnarray}}
\newcommand{\eea}{\end{eqnarray}}
\newcommand{\bwt}{\begin{widetext}}
\newcommand{\ewt}{\end{widetext}}

\newcommand{\bi}{\begin{itemize}}
\newcommand{\ei}{\end{itemize}}
\newcommand{\ben}{\begin{enumerate}}
\newcommand{\een}{\end{enumerate}}
\newcommand{\bca}{\begin{cases}}
\newcommand{\eca}{\end{cases}}
\newcommand{\bln}{\begin{align}}
\newcommand{\eln}{\end{align}}
\newcommand{\bst}{\begin{split}}
\newcommand{\est}{\end{split}}

\def\sigmam{{\sigma^-}}
\def\sigmap{{\sigma^+}}

\begin{document}

\comment{

%\begin{titlepage}

  \begin{center}

\centerline{\Large \bf {On compressing sinh-Gordon solutions}}

\bigskip
\bigskip

{\bf David Vegh}

\bigskip

\small{
{ \it  Centre for Research in String Theory, School of Physics and Astronomy \\
Queen Mary University of London, 327 Mile End Road, London E1 4NS, UK}}

\medskip

{\it email:} \texttt{d.vegh@qmul.ac.uk}

\medskip

{\it \today}

\bigskip

%\date{\today}

}

\preprint{QMUL-PH-21-13}

\title{On compressing sinh-Gordon solutions}

\author{David Vegh}
\email{d.vegh@qmul.ac.uk}

\affiliation{\it  Centre for Research in String Theory, School of Physics and Astronomy,
Queen Mary University of London, 327 Mile End Road, London E1 4NS, UK}

\begin{abstract}

%approximate auto-B\" acklund

This paper is concerned with a class of approximate non-linear transformations that compress solutions of the (generalized) sinh-Gordon equation into parametrically small regions in two-dimensional spacetime. Given the sinh-Gordon field near a time-slice, a long Nambu-Goto string can be constructed in three-dimensional anti-de Sitter space. The string is then approximated to arbitrary accuracy by a slightly smoothed piecewise linear string of $N$ segments.
%In a given scheme, the smoothing is controlled by a single parameter $\varepsilon$. The induced scalar curvature of the new embedding is concentrated in the vicinity of distinct points on the worldsheet. The corresponding sinh-Gordon field has a comb-like structure and its size is controlled by $\varepsilon$.
%collapses into a point when no smoothing is applied.
The corresponding sinh-Gordon field has a comb-like structure and its size is controlled by the amount of smoothing applied to the segmented string.
In a (singular) large-$N$ limit, the transformation commutes with time evolution. As an example, a static cosh-Gordon solution is discussed in detail.
The corresponding smooth and segmented string solutions are obtained and the compressed cosh-Gordon potential is investigated. %Finally, a potential application of the transformation is discussed.

% in the sinh-Gordon theory.

%The transformation approximately commutes with time evolution in the sinh-Gordon theory which becomes exact in the (singular) large-$N$ limit.

%In the large-$N$ limit, the transformation commutes with time evolution in the sinh-Gordon theory.

%The transformation commutes with time evolution in the sinh-Gordon theory in the large-$N$ limit.

\end{abstract}
%
%\end{center}

%\end{titlepage}

\maketitle

%\section{Introduction}
\noindent \textbf{1. Introduction.}
Completely integrable models in two dimensions play an important role in physics. Since they possess sufficiently many conserved quantities they are exactly solvable thus providing a controlled laboratory in which various phenomena can be studied. The focus of this paper is on the hyperbolic version  of the celebrated sine-Gordon equation: the  classical (generalized) sinh-Gordon model. The equation of motion is given by
\be
   \label{eq:sg}
   \p_+ \p_- \alpha + e^{\alpha} - u v e^{-\alpha} = 0
\ee
where $\alpha(\sigma^+, \sigma^-)$ is the sinh-Gordon field, $u(\sigmam)$ and $v(\sigmap)$ are auxiliary fields, and $\sigma^\pm = \half(\tau \pm \sigma)$ are lightcone coordinates in two dimensions ($\p_- \equiv \p_\sigmam$ and $\p_+ \equiv \p_\sigmap$).
The equation describes constant mean curvature surfaces in $\RR^{2,1}$, extremal surfaces in AdS$_3$ \cite{Pohlmeyer:1975nb}, and it governs the evolution of the mesonic mean field in the two-dimensional Gross-Neveu model \cite{Gross:1974jv, Neveu:1977cr}.

In regions where $uv=0$, eqn. (\ref{eq:sg}) reduces to the Liouville equation.
Outside these regions, after a coordinate transformation given by $d\tilde\sigma^-=\sqrt{|u(\sigmam)|}d\sigma^-$,  $d\tilde\sigma^+=\sqrt{|v(\sigmap)|}d\sigmap$, the equation takes the standard sinh-Gordon or cosh-Gordon form
\be
   \tilde\p_+ \tilde\p_- \tilde\alpha + e^{\tilde\alpha} \pm e^{-\tilde\alpha} = 0
\ee
where $\tilde\alpha = \alpha - \log \sqrt{|uv|}$. The variables $\tilde\sigma^\pm$ will be called {\it balanced coordinates}.

The non-linear sinh-Gordon equation is integrable \cite{Pohlmeyer:1975nb, DeVega:1992xc} and possesses singular soliton solutions. Multi-soliton solutions can be constructed by the inverse scattering method.
In this Letter the relationship between the sinh-Gordon equation and long Nambu-Goto strings (extremal surfaces) in AdS$_3$ will be exploited to obtain scheme-dependent non-linear transformations
which compress solution of the sinh-Gordon equation   into parametrically small regions. The flowchart of the map is illustrated in FIG.\ref{fig:flow}.

%similarly to the sine-Gordon equation. The generalized sinh-Gordon equation describes the time evolution of the mesonic mean field in the 2d Gross-Neveu model. It is also related to the motion of Nambu-Goto strings, which will be the topic of the next section.

\begin{figure}[h]
\begin{center}
\includegraphics[width=7.6cm]{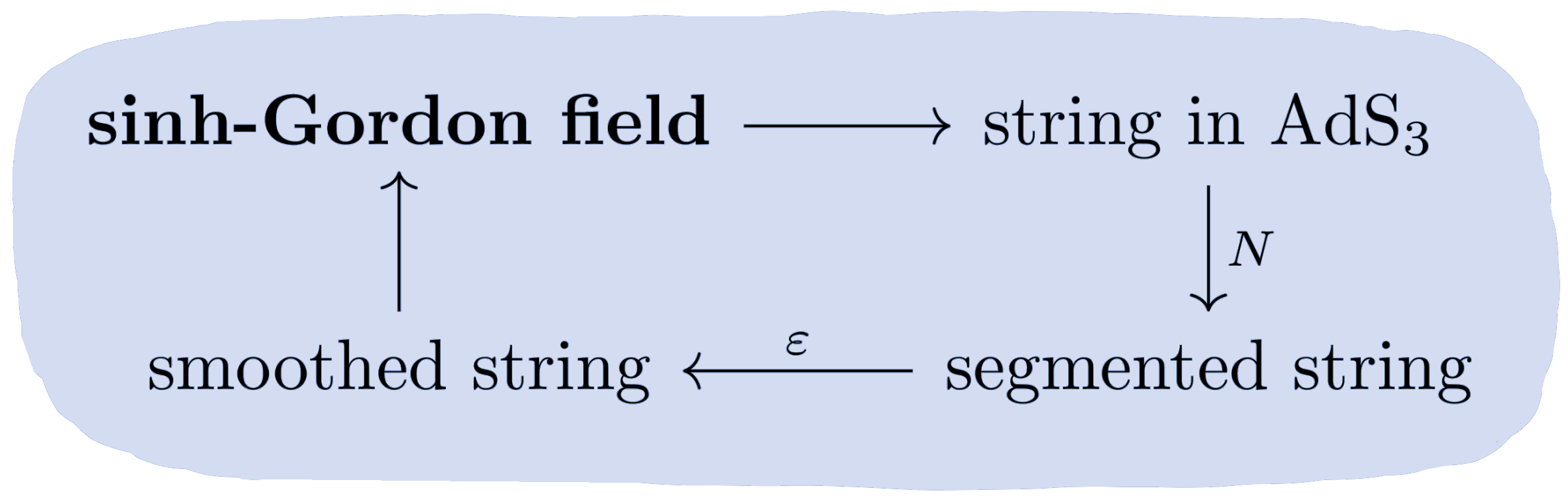}
\caption{\label{fig:flow}  Flowchart of the transformation. In a given segmenting and smoothing scheme the transformed field will depend on two control parameters: the number of string segments $N$ and a smoothing parameter $\varepsilon$.
}
\end{center}
\end{figure}

\comment{

\begin{figure}[h]
\begin{center}
{
\normalsize  %\textrm{scattering}
\[ \begin{tikzcd}
\textrm{\bf sinh-Gordon field} \arrow{r}{}   & \textrm{string in AdS$_3$} \arrow{d}{N} \\
\textrm{smoothed string} \arrow[u] & \textrm{segmented string} \arrow[swap]{l}{\varepsilon} %\\
%\textrm{new sinh-Gordon field}   &
\end{tikzcd}
\]
%\[ \begin{tikzcd}
% & \textrm{string in AdS$_3$}  \arrow{ddr}{N}  & \\
%\textrm{sinh-Gordon field} \arrow{ur}    &  & \\
%& & \textrm{segmented string} \arrow{ddl}{\varepsilon} \\
%\textrm{new shG field}   & & \\
%& \textrm{smoothed string} \arrow{ul}  &
%\end{tikzcd}
%\]
}
\caption{\label{fig:flow}  The flowchart of the compressing transformation. In a given segmentation and smoothing scheme the transformed field will depend on two control parameters: the number of string segments $N$ and a smoothing parameter $\varepsilon$.
}
\end{center}
\end{figure}
}

\noindent

%Starting from a solution of the sinh-Gordon equation, a string embedding will be constructed.  The smooth string is approximated to by a segmented string. Segmented strings are piecewise linear strings consisting of $N$ elementary AdS$_2$ segments.

%Kinks between the segments move with the speed of light which ensures that the string remains segmented at all times. The scalar curvature of the induced metric is constant everywhere, except for points on the worldsheet where left- and right-moving kinks collide where it diverges (similarly to the case of a polyhedron). In order to eliminate these divergences, the segmented string is slightly smoothed in section 4. From the induced metric of the smoothed string a new  sinh-Gordon field can be computed which occupies a smaller region in the two-dimensional ($\tau, \sigma$) spacetime. After giving a detailed example for the transformation in section 5, we discuss a potential application in section 6.

\noindent \textbf{2. Strings in AdS$_3$.} In the first step of the transformation one needs to obtain a smooth string embedding in AdS$_3$.
A unit size AdS$_3$ can be immersed into an $\RR^{2,2}$ ambient space via the equation,
\be
  \nonumber
%  \label{eq:hyp}
   Y \cdot  Y \equiv -Y_{-1}^2 - Y_0^2 + Y_1^2 + Y_2^2 = -1 \qquad  Y \in \RR^{2,2}.
\ee
A part of global AdS$_3$ is covered by the \poincare patch.
The metric on the patch is given by
\be
  \nonumber
  ds^2 = {-dt^2 + dx^2 + dy^2 \over y^2} .
\ee
The coordinates $t, x, y$ are related to $ Y$ via the following transformation
\be
  \nonumber
  (t, \, x, \, y) =
  \le(  {Y_{0} \over Y_{2} - Y_{-1}}, \ {Y_{1} \over Y_{2} - Y_{-1}}, \ {1 \over Y_{2} - Y_{-1}} \ri),
\ee
whose inverse on the hyperboloid is
\be
  \nonumber
   Y = \le( {-1+ t^2 -x^2 - y^2 \ov 2y}, \, {t\ov y} ,  \,{x\ov y} ,  \, {1+ t^2 -x^2 - y^2  \ov 2y}\ri).
\ee
The boundary of AdS is located at $y=0$. In the ambient space the boundary is the set of points that satisfy $ Y^2 = 0$ with the identification $ Y \cong a  Y$ (with $a \in \RR^+$).

The string can be mapped into AdS$_3$ by taking the target space to be $\RR^{2,2}$ and then forcing the string to lie on the hyperboloid using a Lagrange multiplier. In conformal gauge the action is given by
\be
   \nonumber
 S  = -{T \ov 2}\int d\tau d\sigma ( \p_\sigma Y^\mu \p_\sigma Y_\mu - \p_\tau Y^\mu \p_\tau Y_\mu + \lambda( Y^2 + 1)) ,
\ee
where $T$ is the string tension and $ Y(\tau, \sigma) \in \RR^{2,2}$ is the embedding function.
%In terms of the $\vec Y \in \RR^{2,2}$ coordinates, the resulting
The equation of motion in lightcone coordinates is
\be
   \nonumber
 \label{eq:eoms}
  \p_+ \p_- Y - (\p_+ Y \cdot \p_- Y ) Y = 0 \, .
\ee
Due to the gauge choice, the equations are supplemented by the Virasoro constraints
\be
  \nonumber
  \p_- Y \cdot \p_- Y = \p_+  Y \cdot \p_+ Y = 0 \, .
\ee
From the string embedding the sinh-Gordon field can be computed via
\be
  \alpha = \log| \p_- Y \cdot \p_+ Y | \, .
\ee
%\be
%  \beta \equiv \log \p_- N \cdot \p_+ N
%\ee
If the unit normal vector $N$ is defined such that
\be
  \nonumber
    N_a  = e^{-\alpha} \epsilon_{abcd} Y^b \p_- Y^c \p_+ Y^d
\ee
then the auxiliary fields are given by
\bea
  \nonumber
  u &=& -N\cdot \p_-\p_- Y  = \p_- N \cdot\p_- Y = - Y \cdot \p_-\p_- N \\
%  u &=& N\cdot \p_-\p_- Y  = - \p_- N \cdot\p_- Y =  Y \cdot \p_-\p_- N \\
  \label{eq:discv}
  v &=&  N\cdot \p_+\p_+ Y  = -\p_+ N \cdot\p_+ Y =  Y \cdot \p_+\p_+ N
\eea
The string equations of motion guarantee that $u = u(\sigmam)$ and $v=v(\sigmap)$ and that the quantities satisfy the generalized sinh-Gordon equation \cite{Pohlmeyer:1975nb, DeVega:1992xc}. Note that while changing $N \to -N$ flips the signs of both $u$ and $v$, their product which appears in (\ref{eq:sg}) remains invariant.
%\be
%  \p_- Y \cdot \p_+ N = \p_+ Y \cdot \p_- N =0
%\ee

\vskip 0.3cm

\noindent \textbf{3. Auxiliary linear system.} Given a solution of the generalized sinh-Gordon equation, one would like to construct the string embedding. In order to do this one has to solve an auxilliary scattering problem. In terms of $\alpha, u, v$ the $SL(2)$ Lax matrices are  \cite{Alday:2009yn}
\bea
\nonumber
B^-_L=\begin{pmatrix}
{1 \over 4} \p_- \alpha & {1\ov \sqrt{2}} {e^{ {\alpha \ov 2}}  }\cr
{u\ov \sqrt{2}}{e^{-  {\alpha \ov 2}}  }  & -{1 \over 4} \p_- \alpha
\end{pmatrix} &&
B^+_L=\begin{pmatrix}
-{1 \over 4} \p_+ \alpha & -{v \ov \sqrt{2}}{e^{-{\alpha \ov 2} }   }\cr
-{1\ov \sqrt{2}} e^{ {\alpha \ov 2}}    & {1 \over 4} \p_+ \alpha
\end{pmatrix} \\
\nonumber
B^-_R=\begin{pmatrix}
-{1 \over 4} \p_- \alpha &{-{u\ov \sqrt{2}} e^{- {\alpha \ov 2}}   }   \cr
{1\ov \sqrt{2}}{e^{ {\alpha \ov 2}}  } & {1 \over 4} \p_- \alpha
\end{pmatrix}
&&
B^+_R=\begin{pmatrix}{1 \over 4} \p_+ \alpha & -{1\ov \sqrt{2}}{e^{ {\alpha \ov 2}} }  \cr
{v\ov \sqrt{2}}{e^{- {\alpha \ov 2}}   }  & -{1 \over 4} \p_+ \alpha \end{pmatrix} \, .
\eea
It is easy to check that the flatness conditions
\bea
\nonumber
\p_- B^+_L-\p_+ B^-_L+[B^-_L,B^+_L] &=& 0 \\
\nonumber
\p_- B^+_R-\p_+ B^-_R+[B^-_R,B^+_R] &=& 0
\eea
are in fact equivalent to (\ref{eq:sg}).
Consider the left and right auxiliary linear systems
\bea
\nonumber
\p_- \psi^L_{\alpha}+(B^-_L)_{\alpha}^{~\beta}\psi^L_{\beta}=0 \, , & &
\p_+ \psi^L_{\alpha}+(B^+_L)_{\alpha}^{~\beta}\psi^L_{\beta}=0 \, , \\
\label{eq:linear}
\p_- \psi^R_{\dot \alpha }+(B^-_R)_{\dot \alpha}^{~\dot \beta}\psi^R_{\dot \beta } =0 \, , & &
\p_+ \psi^R_{\dot \alpha }+(B^+_R)_{\dot \alpha}^{~\dot \beta}\psi^R_{\dot \beta } =0 \, . \quad
\eea

\begin{figure}[h]
\begin{center}
\includegraphics[width=5.0cm]{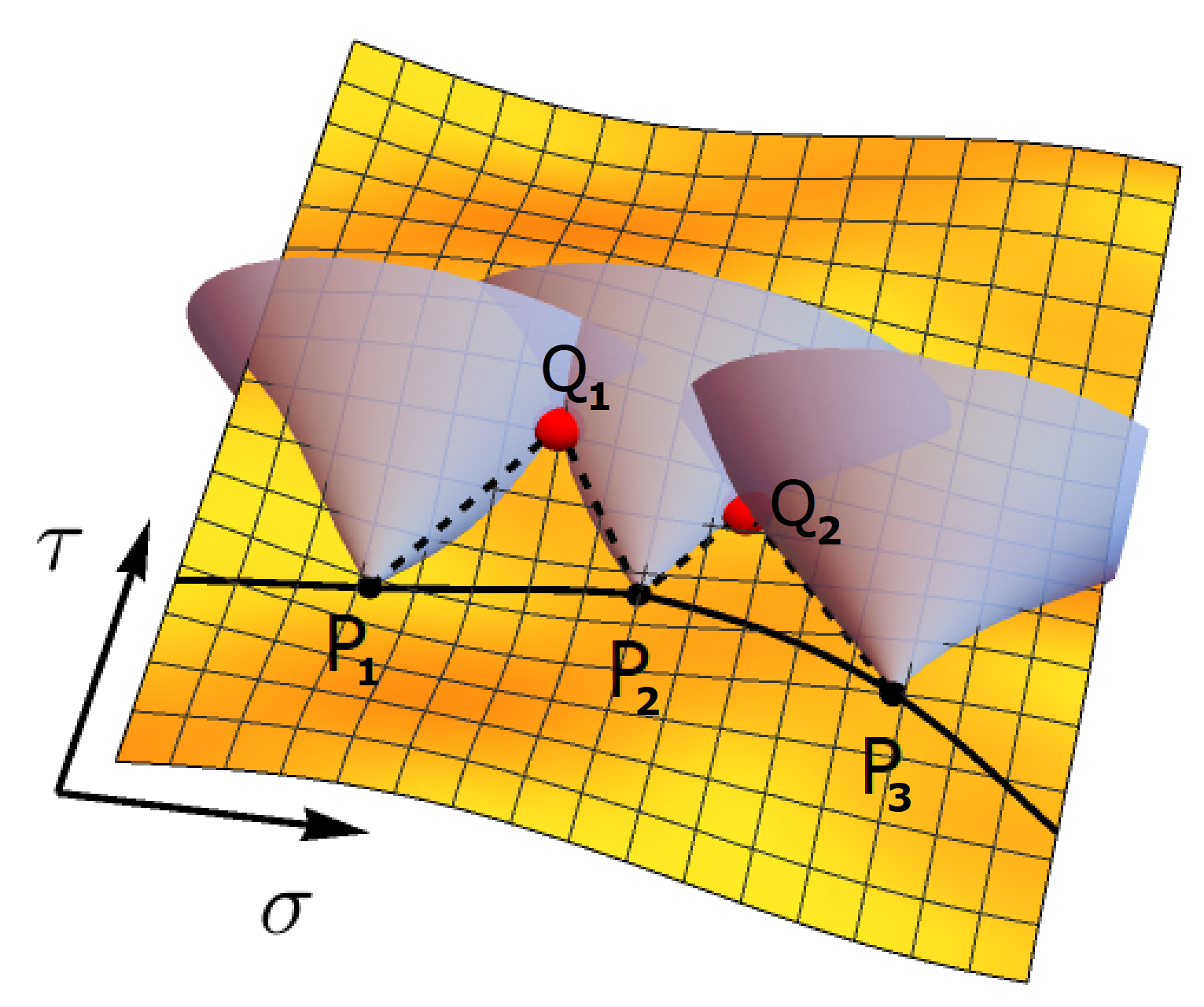}
\caption{\label{fig:getp} Approximating the smooth embedding (yellow surface) with a segmented string. After discretizing the string on a time-slice (three black dots on a line) one can determine future null-separated points on the worldsheet (two red dots). The resulting zigzag (dashed lines) constitutes the initial condition for the segmented string.
}
\end{center}
\end{figure}

\noindent
where $\psi^L_{\alpha}$ and $\psi^R_{\dot \alpha }$ are two-component spinors\footnote{The index $\alpha$ is not to be confused with the sinh-Gordon potential.}. Each of these systems have two linearly independent solutions, denoted by $\psi^L_{\alpha, a}$  and $\psi^R_{\dot \alpha, \dot a}$, (with $a,\dot a, \alpha, \dot\alpha =1,2)$.
These can be normalized so that
\be
  \epsilon^{\beta \alpha } \psi^L_{\alpha,a} \psi^L_{\beta,b}=\epsilon_{a b},~~~~~\epsilon^{\dot \beta \dot \alpha } \psi^R_{\dot \alpha, \dot a} \psi^R_{\dot \beta,\dot b}=\epsilon_{\dot a \dot b}
\ee
where $\epsilon$ is the $2\times 2$ Levi-Civita tensor.
Finally, the string embedding is given by\footnote{For the scattering problem in an $SU(1,1)$ basis, see \cite{Jevicki:2007aa}.}
\be
 Y_{a \dot a } \equiv
\begin{pmatrix}
Y_{-1}+Y_{2}& Y_1-Y_0\cr Y_1 + Y_0 & Y_{-1}-Y_{2}
\end{pmatrix}_{a,\dot{a}}=
\psi^L_{\alpha,a} M^{\alpha \dot \beta} \psi^R_{\dot \beta ,\dot{a}}
\ee
where $M = \textrm{diag}(1,1)$.

\vskip 0.3cm

\noindent \textbf{4. Segmented strings.} Once the smooth\footnote{In general the embedding will only be smooth almost everywhere. Namely, the string may contain cusps which correspond to singular solitons in the sinh-Gordon theory.} string embedding is obtained, one can proceed by approximating it with a segmented string.
Segmented strings are AdS generalizations of piecewise linear
strings in flat space \cite{Vegh:2015ska, Callebaut:2015fsa}.
They consist of elementary AdS$_2$ segments which themselves are linear subspaces in $\RR^{2,2}$.
Kinks between the segments move with the speed of light which ensures that the string remains segmented at all times. The scalar curvature of the induced metric is constant everywhere, except for points on the worldsheet where left- and right-moving kinks collide. At these locations the curvature diverges (similarly to the case of a polyhedron).

The segmentation procedure is depicted in FIG.~\ref{fig:getp}. The diagram shows the string worldsheet (yellow surface) embedded into AdS$_3$. A time-slice on the worldsheet is indicated by a black line. One can discretize this curve according to any preferred scheme. This will result in a set of points $\{ P_i  \}$.

\begin{figure}[h]
\begin{center}
\includegraphics[width=6cm]{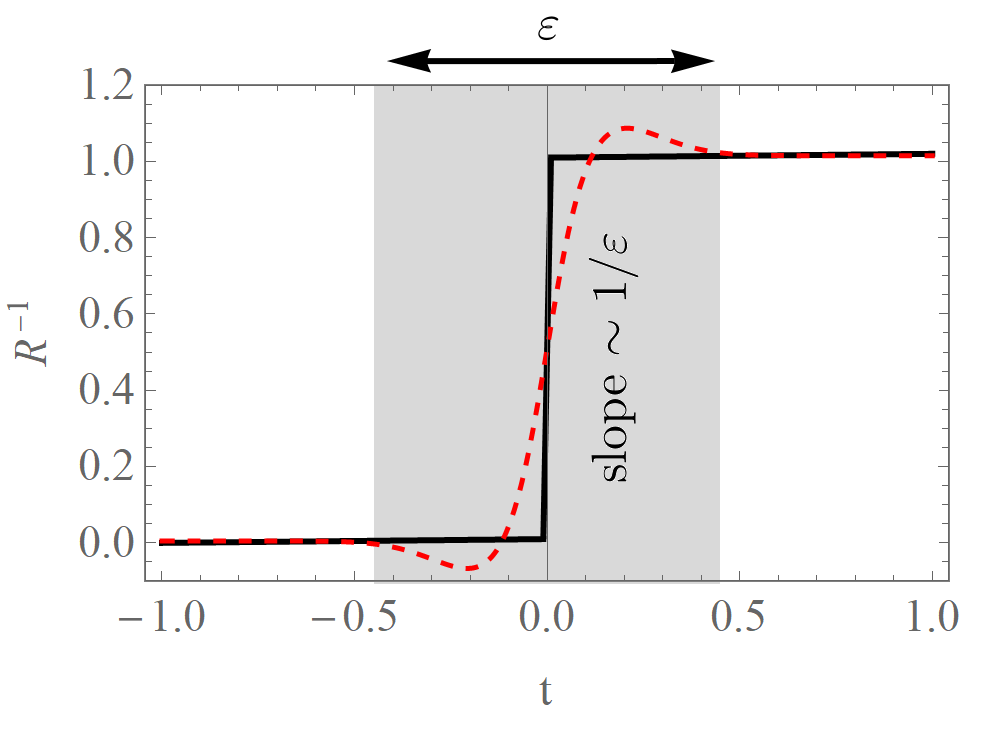}
\caption{\label{fig:smoothing} Adjacent string segments have different inverse radii. The jump can be smoothed by an appropriate function (dashed line) over a region of size $\varepsilon$ (shaded area) so that the parameters of the two AdS$_2$ segments outside the region do not change.
}
\end{center}
\end{figure}

\noindent These points are indicated by black dots in the figure. For a full set of initial data, one needs to further specify another set of points denoted by  $\{ Q_i \}$ (red dots in FIG.~\ref{fig:getp}) which are null-separated from their neighbors.
This condition means that they must lie at the triple intersection of future lightcones and the smooth worldsheet as in the figure. The initial data for the segmented string will then be defined by the null zigzag $P_1, Q_1, P_2, Q_2, \ldots$. If the points are taken to be in $\RR^{2,2}$, then any three consecutive points define a linear AdS$_2$ subspace: a segment of the string\footnote{The segmented string is an exact Nambu-Goto string solution. Time evolution   can be obtained by reflecting the points on their two neighbors (e.g. $P_2$ on $Q_1$ and $Q_2$) as described in  \cite{Vegh:2015ska, Callebaut:2015fsa, Gubser:2016zyw, Vegh:2016hwq,  Vegh:2018dda}.
%The technique is ideally suited for calculations since there are no accumulating numerical errors (see also  \cite{}).
}.
On the \poincare patch the segments are given by the equations
\be
  \label{eq:semic}
 (x-x_i)^2 + y^2 -(t-t_i)^2=R_i^2
\ee
where $i$ labels the segments. Each segment has three parameters $\{ x_i, t_i, R_i \}$.
The total number of segments will be denoted by $N$.

%If the points are taken to be in $\RR^{2,2}$, then they satisfy
%\be
% \nonumber
%  (Q_i - P_i)^2 =  (Q_i - P_{i+1})^2  = 0  \, .
%\ee

\vskip 0.3cm

\noindent \textbf{4. Smoothing.} One can compute the generalized sinh-Gordon fields corresponding to a segmented string. Inside the AdS$_2$ patches $u=v=0$, on the kink worldlines and at the kink collision points (vertices) they are ill-defined.
In order for these fields to take finite values, the segmented string has to be smoothed. This can be achieved by smoothing the initial conditions, for instance by using Mikhailov's solution \cite{Mikhailov:2003er}. The embedding is given in terms of the position function of the string endpoint on the boundary,
\bea
  \nonumber
  t(t_{r}, y) &=& t_{r} + {y\ov \sqrt{1-x_0'(t_{r})^2}} \\
  %\label{eq:mikh}
  \nonumber
  x(t_{r}, y) &=& x_0(t_{r}) + {y \, x_0'(t_{r})\ov \sqrt{1-x_0'(t_{r})^2}}
\eea

\begin{figure}[h]
\begin{center}
\includegraphics[width=6cm]{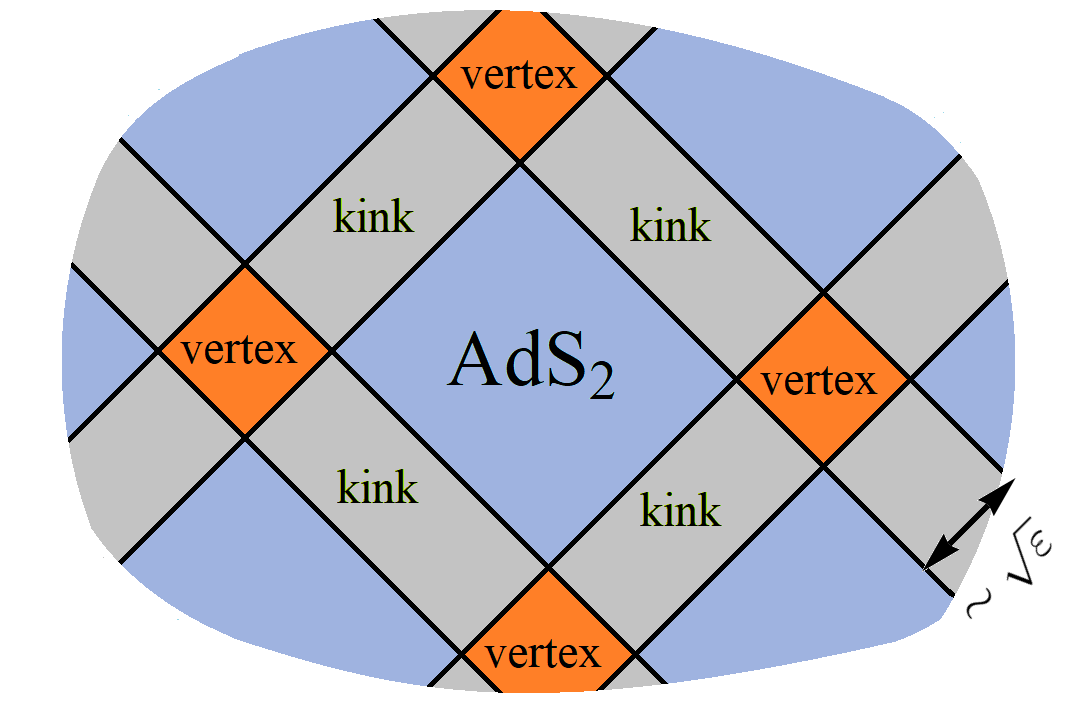}
\caption{\label{fig:wsheet} A patch of the worldsheet of the smoothed segmented string. Kinks between AdS$_2$ string segments (blue regions) occupy finite size regions (gray areas). In balanced coordinates, their linear size is proportional to $\sqrt{\varepsilon}$. If the smoothing does not affect the AdS$_2$ patches, then $u=v=0$ in these regions. Hence in balanced coordinates they are compressed into points, kinks are squeezed into lines, and
the vertex regions (orange diamonds) fill the entire worldsheet.
}
\end{center}
\end{figure}

\noindent where $x_0(t_{r})$ specifies the endpoint of the string in terms of the retarded time $t_{r}$.
The resulting string  will have non-linear waves moving in one direction.

Consider the motion given by
\be
\nonumber
 x_0(t_{r})=\theta(\sqrt{1+t^2}-1) \, ,
\ee
where $\theta$ is the Heaviside function. Mikhailov's embedding gives a segmented string that consists of two pieces: a vertical segment at $x=0$ attached to another one with radius $R=1$. The relativistic acceleration of the string endpoint (the inverse radius of the segment) jumps as seen in FIG.~\ref{fig:smoothing}.
One can smooth the kink by choosing a new $\tilde x_0(t_{r})$ function which is equal to $x_0(t_{r})$ outside the region $(-{\varepsilon \ov 2}, {\varepsilon \ov 2})$ and whose acceleration is smooth  (red dashed line in the figure).
The average slope of the acceleration function will be of the order of $\Delta a / \varepsilon$ where $\Delta a$ is the jump in the acceleration. In order to have an estimate for the $u$ field across this region, the smooth motion of the endpoint can be approximated by taking $\tilde x_0(t_{r}) = {\Delta a  t_r^3 /  \varepsilon}$ which will produce a linear acceleration function for small positive $t_r$. After changing to lightcone coordinates in Mikhailov's embedding, one gets $u \approx {6 \Delta a / \varepsilon}$. This means that in balanced coordinates the width of the smoothed kink is (see FIG.~\ref{fig:wsheet})
\be
  \nonumber
L \propto \sqrt{u} \,  \varepsilon \propto \sqrt{\Delta a \varepsilon} \, .
\ee
If there are $N$ segments, then $\Delta a$ scales as $N^{-1}$ and $\varepsilon$ can also be taken to scale in the same way. The balanced linear size on the worldsheet is then proportional to $N^0 \sqrt{\varepsilon}$.

In the final step of the transformation, the sinh-Gordon fields $\alpha, u, v$ are computed. As $N \to \infty$, the segmented string approaches the original smooth string and thus their time evolution will be the same.

\clearpage

\begin{figure}[h]
\begin{center}
\includegraphics[width=6cm]{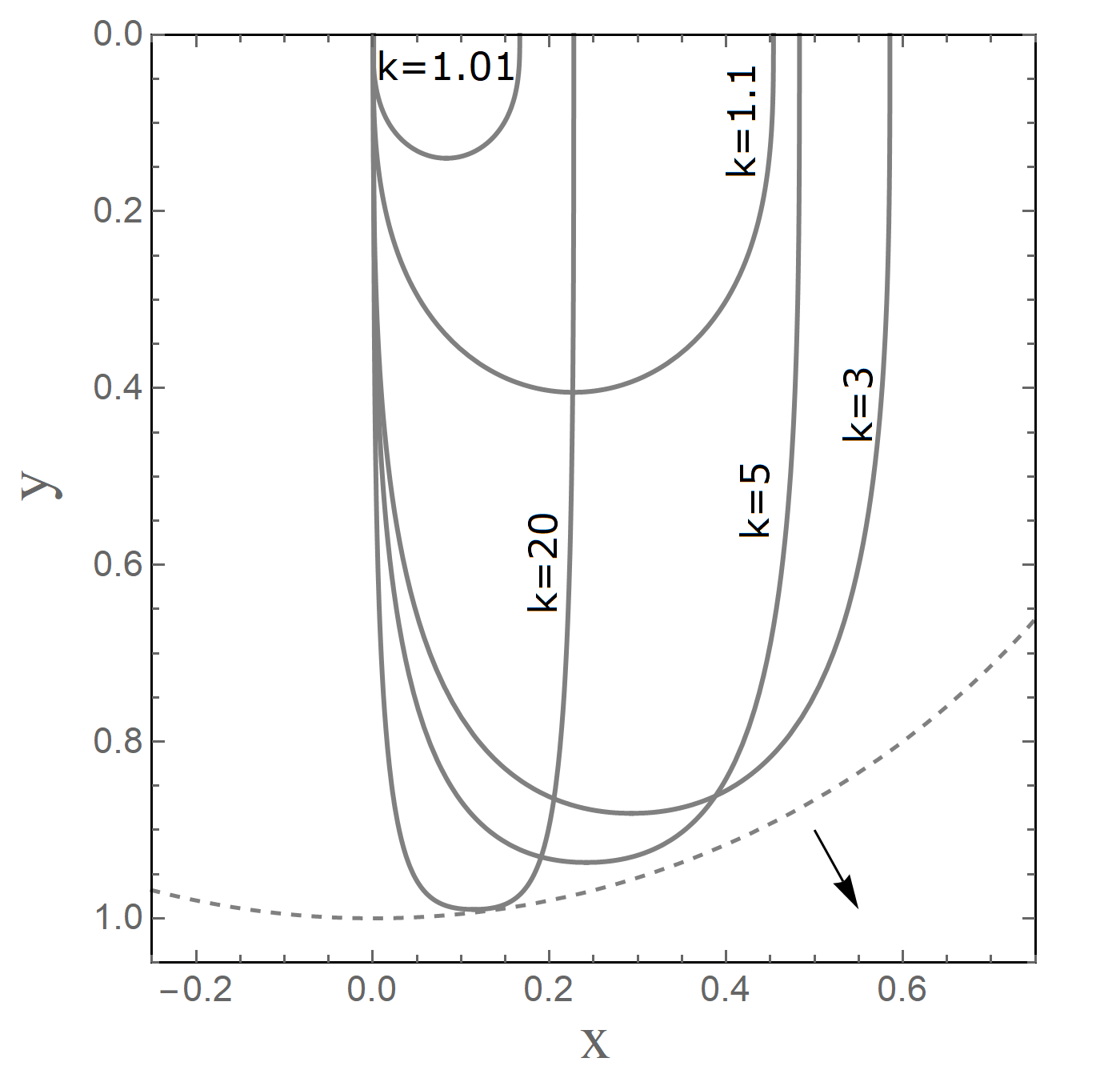}
\caption{\label{fig:kgen} Scaling string embeddings for various values for the parameter $k$. The figure corresponds to $t=1$ on the \poincare patch.  In these coordinates, the size of the string increases linearly with time, but its shape does not change. The dashed line indicates the lightcone emanating from the event at $x=y=t=0$.
}
\end{center}
\end{figure}

\comment{
\begin{figure}[h]
\begin{center}
\includegraphics[width=8cm]{fig_segmented3b.png}
\caption{\label{fig:segm}  Segmented string (with $N=13$ segments) approximating a static smooth embedding (gray curve in the background). Kinks move with the speed of light in the direction indicated by the arrows.
}
\end{center}
\end{figure}

\begin{figure}[h]
\begin{center}
\includegraphics[width=8cm]{fig_segmented4.png}
\caption{\label{fig:smooth}  Smoothed segmented string.
}
\end{center}
\end{figure}
}

%\section{An example: string hanging from the boundary}
\noindent \textbf{5. An example.} Consider the function
\be
  \label{eq:stata}
  \alpha = -\log \le[ k \, \textrm{sn}^2\le( {\sigma \over \sqrt{2k}}| -k^2 \ri) \ri]
\ee
where sn$(u|  m)$ is the Jacobi elliptic function and $k\ge 1$ is a parameter. It solves (\ref{eq:sg}) with $u=-v=1$, i.e. it is a static cosh-Gordon solution.

In order to compress the function, the first step is constructing the corresponding string embedding. The spinor solutions of (\ref{eq:linear}) are found to be
{\footnotesize
\bea
   \nonumber
 \psi^L_{1} &=&   e^{\om_-}
   \le( \sqrt{\frac{2}{\text{sn}}+2 k \, \text{sn}} , \  -\frac{\sqrt{2} \left(\text{cn} \, \text{dn}+\text{sn}\sqrt{k^2-1} \right)}{\sqrt{\text{sn}+k
   \text{sn}^3}} \ri) \\
   \nonumber
 \psi^L_{2} &=&   e^{-\om_-}
   \le( {\sqrt{\frac{1+ k \text{sn}^2}{8\text{sn} \left(k^2-1\right) }}}  , \ \frac{\text{sn}\sqrt{k^2-1} -\text{cn} \,
   \text{dn}}{  \sqrt{8\left(k^2-1\right)  \left(\text{sn}+ k \text{sn}^3\right)}} \ri) \\
   \nonumber
 \psi^R_{1} &=&   e^{\om_+}
    \le(  -\frac{\sqrt{2} \left(\text{cn} \, \text{dn}+\text{sn}\sqrt{k^2-1} \right)}{\sqrt{\text{sn} - k   \text{sn}^3}}
    , \
   \sqrt{\frac{2}{\text{sn}} -2 k \, \text{sn}}
    \ri) \\
   \nonumber
 \psi^R_{2} &=&   e^{-\om_+}
    \le(
    \frac{-\text{sn}\sqrt{k^2-1} +\text{cn} \,
   \text{dn}}{  \sqrt{8\left(k^2-1\right)  \left(\text{sn}- k \text{sn}^3\right)}}  , \
    - {\sqrt{\frac{1- k \text{sn}^2}{8\left(k^2-1\right)\text{sn}  }}}
  \ri)
\eea
}
where
\bea
\nonumber
\Pi_\pm &\equiv& \Pi \left(\pm k;\text{am}\left(\frac{\sigma }{\sqrt{2k} }|-k^2\right)|-k^2\right) \\
\nonumber
 \om_\pm  &\equiv& \sqrt{k-k^{-1}} \le( -\sqrt{k} \Pi_\pm + {\tau + \sigma \over 2\sqrt{2}} \ri)
\eea
and  sn, cn and dn denote Jacobi elliptic functions with arguments $\left(\frac{\sigma }{\sqrt{2k}}|-k^2\right)$, and am is the Jacobi amplitude function.
If one eliminates $\tau$ in favor of \poincare time $t$, then the embeddings will have the form $x(\sigma, t) = t \, \tilde x(\sigma)$ and $  y(\sigma, t) = t \, \tilde y(\sigma)$. Thus they scale linearly with time while their shape is constant.
%\be
%  \nonumber
%  x(\sigma, t) = t \, \tilde x(\sigma) \, , \qquad  y(\sigma, t) = t \, \tilde y(\sigma) \, .
%\ee

\begin{figure}[h]
\begin{center}
\includegraphics[width=7.5cm]{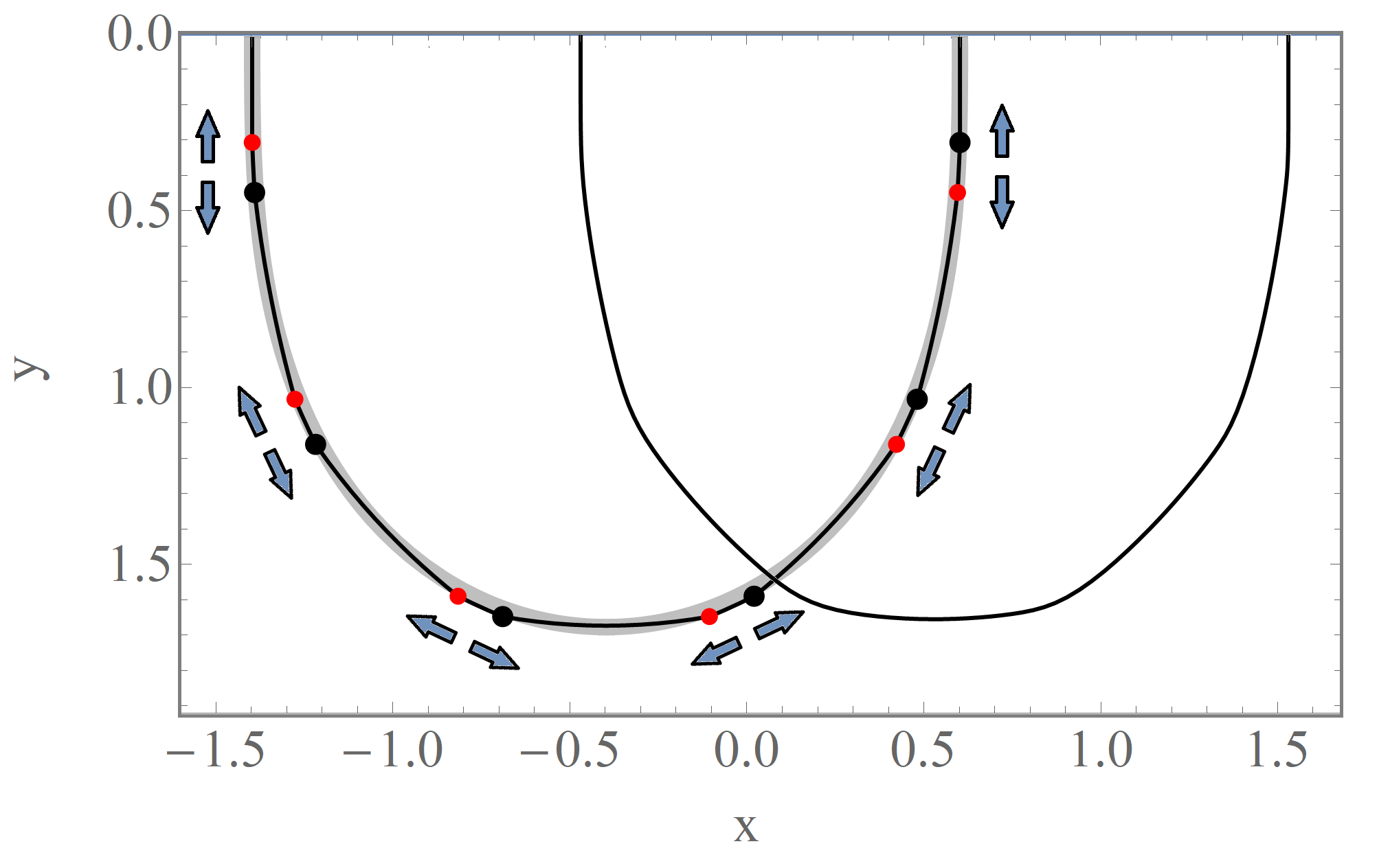}
\caption{\label{fig:kgen}  Segmented string (with $N=13$ segments) approximating a static smooth embedding (gray curve in the background) on the \poincare patch. Kinks move with the speed of light in the direction indicated by the arrows.   Smoothed segmented string.
}
\end{center}
\end{figure}

The  shape $\tilde x(\tilde y)$  of such an embedding satisfies the equation of motion
\be
  \nonumber
\tilde y \left(\tilde x^2+\tilde y^2-1\right) \tilde x''+ \left(2-2 \tilde y^2\right) \tilde x'^3+4 \tilde x \tilde y \tilde x'^2+\left(2-2 \tilde x^2\right) \tilde x'=0
\ee
which can be derived from the Nambu-Goto action. FIG.~\ref{fig:kgen} (left) shows the embedding on the \poincare patch for various values of $k$. As $k\to 1^+$, at $t=1$ the string is mapped into a vanishingly small region near the origin and its size increases slowly as time evolves. By applying an $SO(2,2)$ isometry transformation the size of the string can be kept finite as the limit is taken. Thus  at $k=1$ the string is static. In the following, let us concentrate on this special case.
Its shape $x(y)$  satisfies the equation
\be
  \nonumber
%  2 x'(y)+2 x'(y)^3 - y x''(y) = 0
  2 x'+2 (x')^3 - y x'' = 0
\ee
which is indeed invariant under rescaling the coordinates by a constant factor.
The explicit solution centered at $x=0$ has two branches\footnote{The Ricci scalar of the induced metric is $R = -2 \le(1+ ( y / y_0)^4 \ri)$. Since $R< R_0$ (where $R_0 = -2$ is the curvature of AdS$_3$) the solution in balanced coordinates satisfies the cosh-Gordon equation. Close to the boundary ($y \to 0$) the solution asymptotes to AdS$_2$.}
\be
  \nonumber
  x(y) = \mp x_0 \pm { y^3 \ov 3 y_0^2} \, _2F_1\left({ \frac{1}{2},\frac{3}{4};\frac{7}{4}};  { y^4 \over y_0^4}\right)
\ee
where   $x_0 \equiv \frac{\sqrt{\pi } \Gamma \left(\frac{7}{4}\right)}{3 \Gamma \left(\frac{5}{4}\right)} y_0$ and $y_0$ is a parameter. The string touches the boundary at $x= \pm x_0$.
In FIG.~\ref{fig:kgen} (middle) the embedding is plotted  (background gray curve).

\begin{figure}[h]
\begin{center}
\includegraphics[width=4.0cm]{ 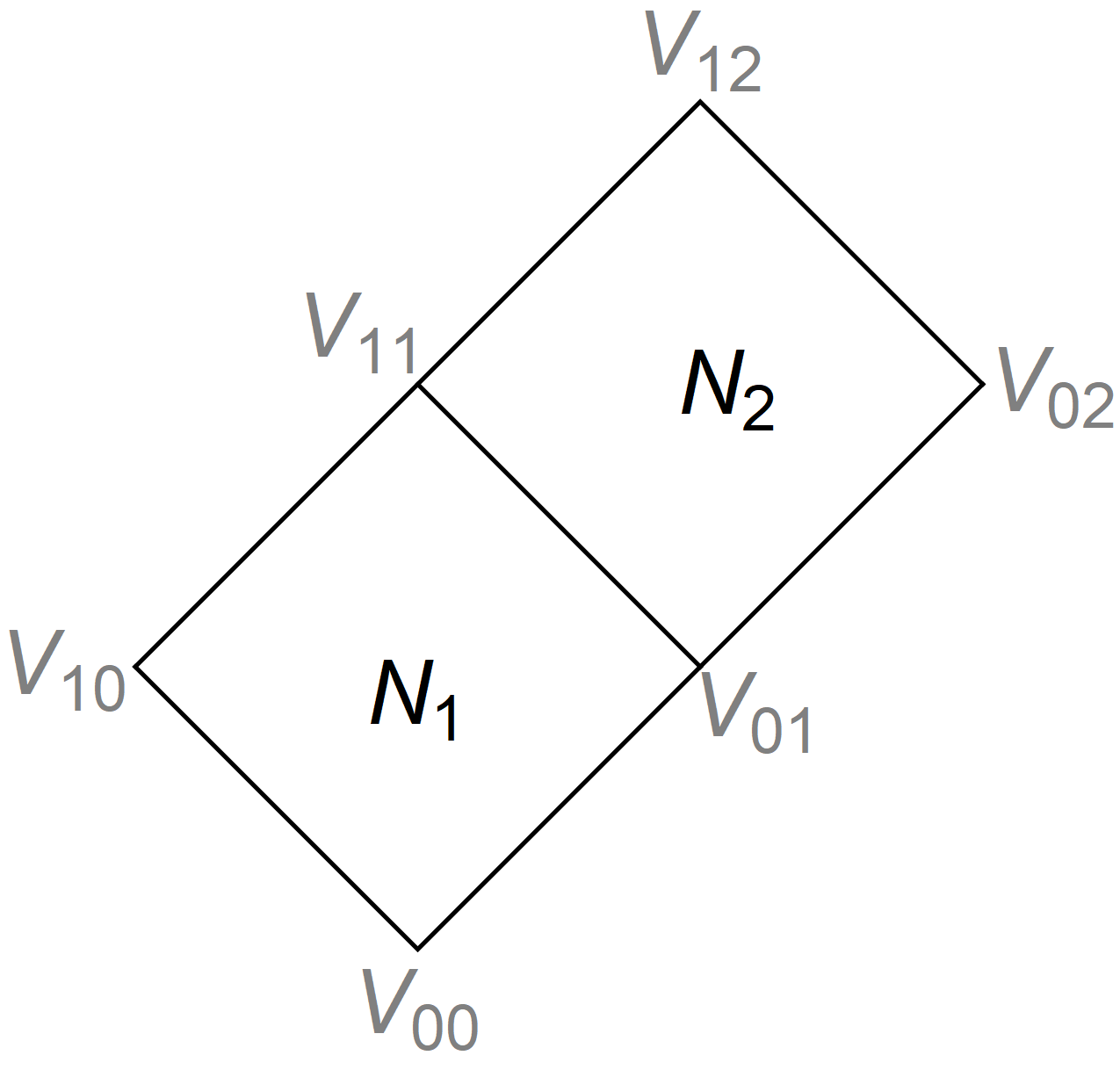}
\caption{\label{fig:twopatches}   Two adjacent AdS$_2$ patches on the worldsheet. Their normal vectors are denoted by $N_1$ and $N_2$.
 }
\end{center}
\end{figure}

\noindent
One can switch to balanced coordinates via the coordinate transformation given by
\be
  \nonumber
  t =  { y_0   \tau \ov \sqrt{2}} \qquad y = y_0 \, \textrm{sn} \le( {\sigma  \over \sqrt{2}}, -1 \ri) \, .
\ee
In these coordinates $u=-v = 1$ and it is  easy to check that $\alpha$ is indeed given by (\ref{eq:stata}) with $k=1$.

The next step is to segment the string. Since in an appropriate frame the string is static, one can apply the shooting technique described in Appendix B of \cite{Vegh:2018dda}. The only parameter is the number $N$ of string segments. The resulting segmented string is periodic in \poincare time. An example with $N=13$ is depicted in FIG.~\ref{fig:kgen} (middle). The black curves are circular arcs given by (\ref{eq:semic}). Red and black dots indicate left- and right-moving kinks between the segments.

The third step is smoothing the segmented string. This can be done in the way described earlier, but in practice it is easier to do the following. One can subdivide the string into many pieces by adding ``null kinks'' \cite{Vegh:2016hwq} which do not further break the individual segments (i.e. they have the same parameters in (\ref{eq:semic})). If each segment is subdivided into $n$ pieces then one has $n N$ segments in total.  After the subdivision, a smoothing filter can be applied on the points.
For instance one can repeatedly apply the transformation
\be
  \nonumber
  P_i' \propto  \lambda (P_{i-1}+ \lambda P_{i+1}) + P_i  \quad \textrm{such that} \quad (P_i')^2 = -1 \, ,
\ee
for points in the refined segmented string, and the same for $Q_i$. This step smoothes the points similarly to a discretized heat equation.
After an iteration $Q'_i - P'_i$ and $Q'_i - P'_{i+1}$ are generically not null, but one can define new $Q''_i$ positions such that they lie in the AdS$_2$ spanned by $\{ P'_i, Q'_i, P'_{i+1} \}$ and that
$(Q''_i - P'_i)^2=(Q''_i - P'_{i+1})^2 =0$, and the points $P''_i = P'_i$ are not changed. In the next iteration one smoothes the initial data given by the null zigzag $\{ P''_i, Q''_i \}$. Let $q$ denote the number of smoothing iterations on the string. For sufficiently large $n$, the resulting string is a good approximation for a smooth string.
An example is depicted in  FIG.~\ref{fig:kgen} (right).

\begin{figure}[h]
\begin{center}
\includegraphics[width=8.0cm]{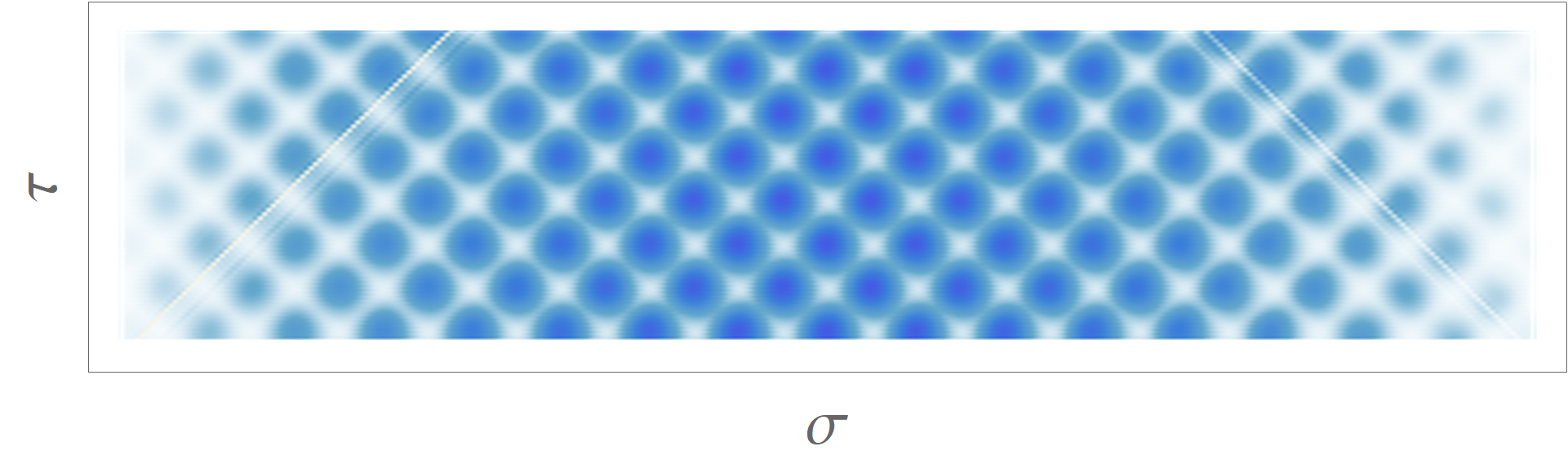}
\includegraphics[width=8.0cm]{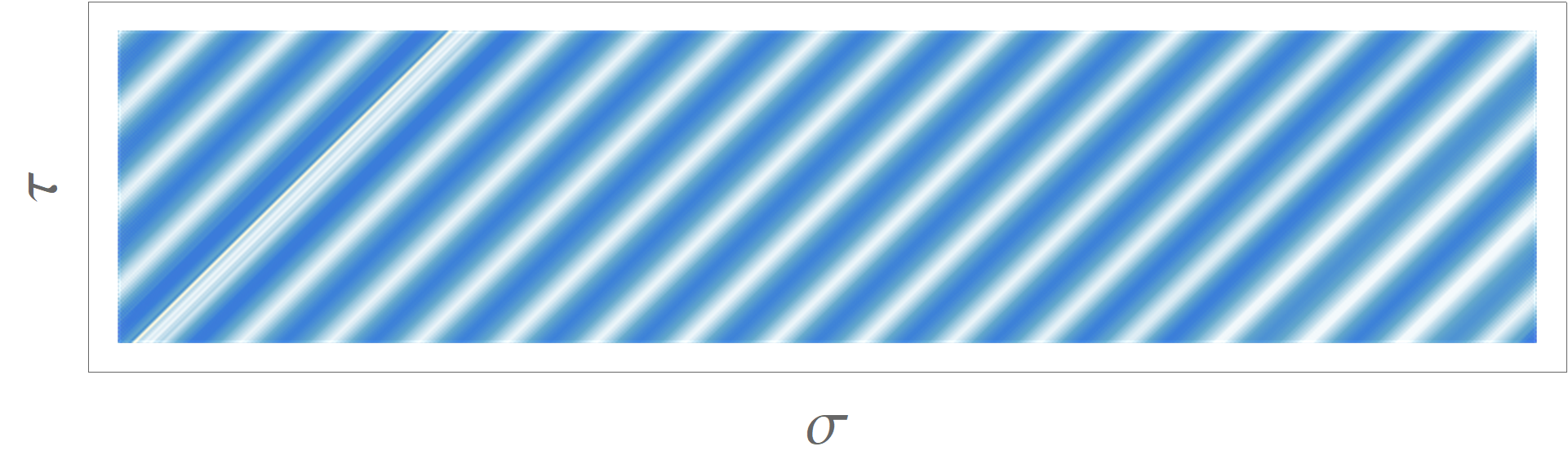}
\includegraphics[width=8cm]{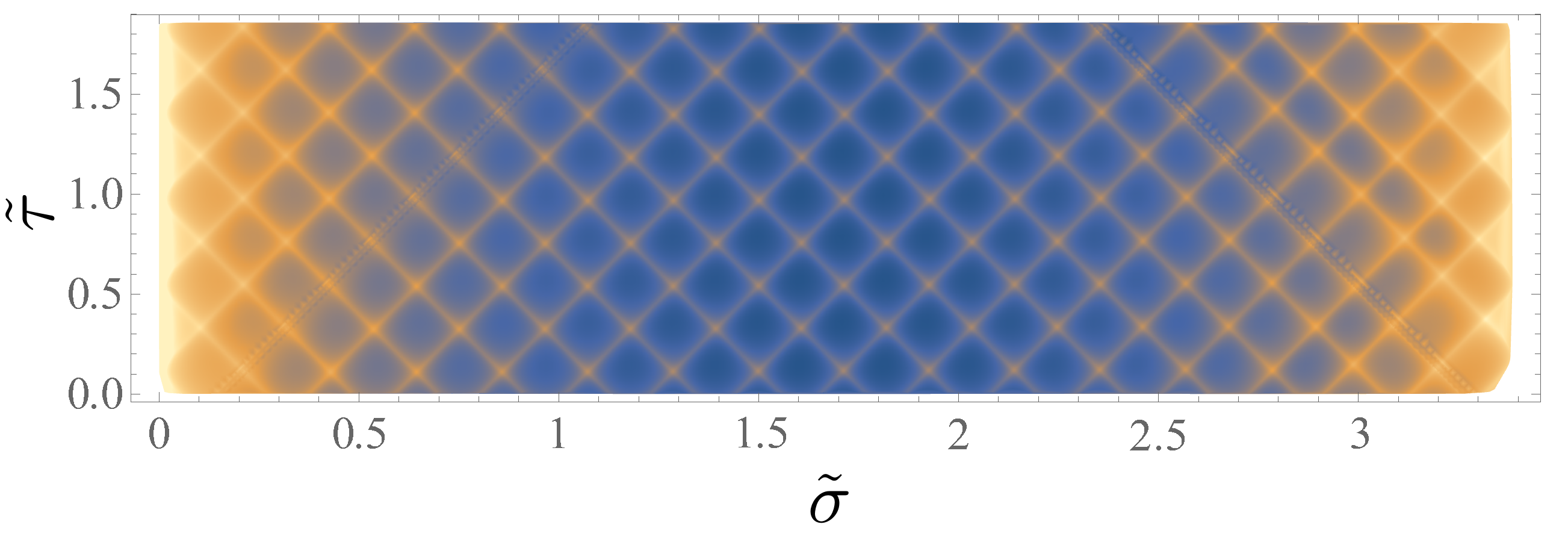}
\caption{\label{fig:ricci} {\it Top:}  Density plot of the Ricci scalar of the induced metric on the  worldsheet of a smoothed segmented string.  The function is approximately periodic in time. {\it Middle:} The $u(\sigma^-)$ function.  {\it Bottom:}  The compressed cosh-Gordon potential $\alpha$ in balanced  $(\tilde\tau, \tilde\sigma)$ coordinates. Peaks move with the speed of light and when they collide the $e^\alpha$ area density becomes large. These points (orange) correspond to the AdS$_2$ patches of the smoothed segmented string. The blue diamonds correspond to smoothed vertices.
}
\end{center}
\end{figure}

Finally, the generalized sinh-Gordon fields $\alpha, u, v$ can be computed using worldsheet points and normal vectors which all take values in $\RR^{2,2}$. For instance in  FIG.~\ref{fig:twopatches} for the patch with normal vector $N_1$ one has
\be
  e^\alpha = a^{-2}(V_{10}-V_{00})(V_{01}-V_{00})
\ee
where $a$ is the lattice spacing. A discrete version of (\ref{eq:discv}) is given by
\be
  v \approx a^{-2}(N_1-N_2)(V_{01}-V_{00})
\ee
and an approximate $u$ function can similarly be computed.
FIG.~\ref{fig:ricci} (middle) shows an example for $v$ computed this way for a slightly smoothed string with $N=30$ segments.
After switching to balanced coordinates, the cosh-Gordon field is seen to have peaks traveling with the speed of light, see FIG.~\ref{fig:ricci} (bottom). Time-slices are show in FIG.~\ref{fig:alpha}. For comparison, the original potential given in (\ref{eq:stata}) with $k=1$ is plotted in black. The difference between the two figures is the number of smoothing iterations applied on the subdivided segmented string. Without any smoothing the balanced width of the string would be zero and the spikes would diverge.

Finally, one can check that the balanced width $L$ of the compressed  function indeed scales as expected. Based on the properties of the heat equation, one expects that an appropriate smoothing parameter can be defined as $\varepsilon = \sqrt{q}/n$ and that $L \propto \sqrt{\varepsilon}$. This is indeed born out by numerical calculations: for an example with $N=10$ segments a numerical fit gave $L \propto n^\gamma$ with $\gamma = -0.495 \pm 0.01$ and $L \propto q^\delta$ with $\delta = 0.246 \pm 0.01$, confirming the expectations.

\clearpage

\begin{figure}[h]
\begin{center}
\includegraphics[width=8cm]{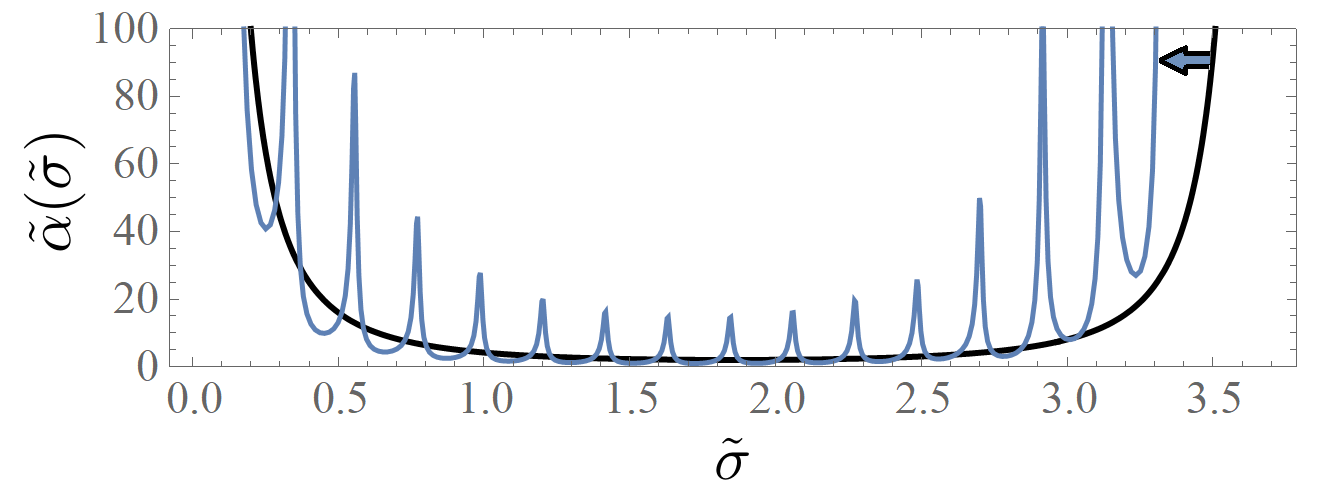}
\includegraphics[width=8cm]{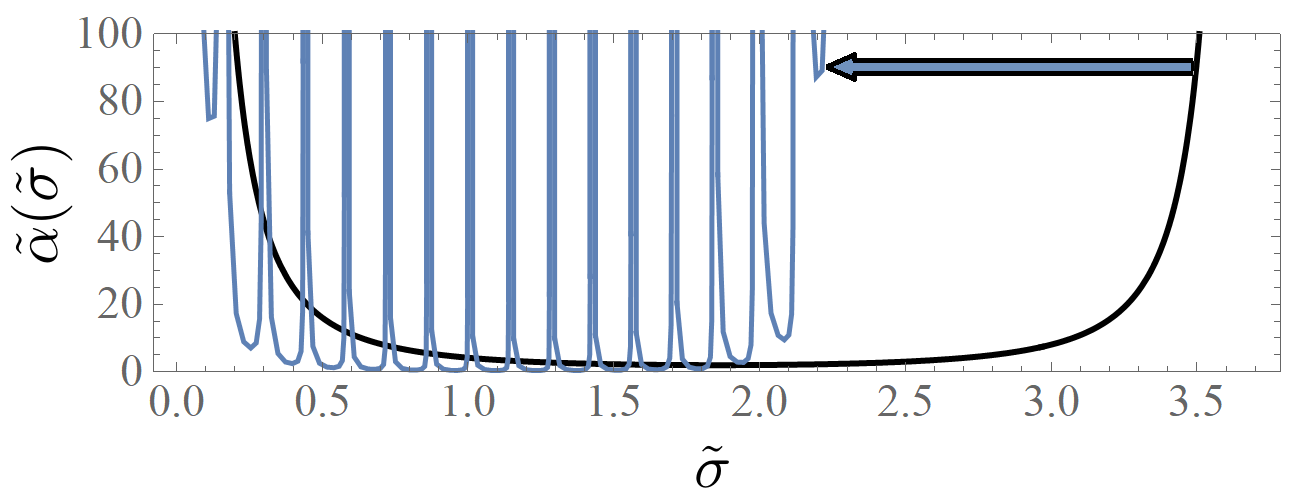}
\caption{\label{fig:alpha} {\it Top:} A static cosh-Gordon solution from eqn. (\ref{eq:stata}) with $k=1$ (black curve) and its compressed version (blue curve) on a time-slice. The compressed function is obtained from a smoothed segmented string with $N=30$ segments. The arrow indicates the reduction of the balanced width. The snapshot is taken at the moment when left- and right-moving peaks collide. {\it Bottom:} A smaller amount of smoothing yields a more compressed (and more singular) potential.
 }
\end{center}
\end{figure}

\noindent \textbf{6. Discussion.} This paper presented a transformation of the generalized sinh-Gordon field that approximately commutes with time-evolution.
The transformation has been defined by constructing from the sinh-Gordon field a long Nambu-Goto string in AdS$_3$. The (almost everywhere) smooth string is approximated by a segmented string. The segmented string is smoothed and the new sinh-Gordon field is computed from the embedding. The new  function
occupies a smaller region in balanced coordinates in which the equation has a standard sinh- or cosh-Gordon form. The transformed function has peaks (see FIG.~\ref{fig:alpha}) which travel with the speed of light.

For the transformation, one can use various segmentation and smoothing  schemes.
In a given scheme one is still left with two parameters: the number of string segments $N$, and the smoothing parameter $\varepsilon$. It is plausible that in an appropriate large-$N$ limit the scheme-dependence goes away. It would be interesting to see if one can define the transformation in a closed form similarly to an auto-B\" acklund transformation.

The equations of motion for a string in AdS$_3$ have an internal $SO(2,2)$ symmetry acting on the vectors  $\{ Y, e^{-{\alpha \ov 2} } \p_- Y, e^{-{\alpha \ov 2} } \p_+ Y, N \}$ \cite{Alday:2009yn}. Note that this symmetry has been broken by both smoothing procedures discussed in the paper.

%\DV{for a clsoed string keep original segmented string points fixed so non-trivial limit at large $\varepsilon$}

%\DV{oversmoothing?}

%\DV{smoothing breaks X/N symmetry}

The results made use of the equivalence of the sinh-Gordon equation and the string equation of motion in AdS$_3$. Similar equivalences exist in higher dimensional AdS spacetimes which allows for the extension of the compressing transformation to other systems, such as the $B_2$ Toda theory which is related to strings moving in AdS$_4$ \cite{Jevicki:2009bv}.

The transformed function contains $N$ peaks which superficially look similar to boosted sinh-Gordon solitons.
Solitons can be thought of as particles whose motion is governed by the hyperbolic Ruijsenaars-Schneider model with a special coupling value \cite{Ruijsenaars:1986vq}. Since the balanced size of the transformed domain vanishes as $\varepsilon \to 0$, it is plausible to think that this gives an interesting limit of the model. Perhaps this limit leads to a dual description of the string.

\noindent \emph{Acknowledgments.}
This work is supported by the STFC Ernest Rutherford grant ST/P004334/1.

\bibliography{broken}

\end{document}